\documentstyle[aps]{revtex}
\begin{document}
\title{ Thermodynamics of multi-boson phenomena }
\author{ Yu.M. Sinyukov\thanks{%
e-mail: sinyukov@bitp.kiev.ua}, S.V. Akkelin }
\address{ Bogolyubov Institute for Theoretical Physics,
\\ Kiev 252143, Metrologicheskaya 14b, Ukraine }
\author{R. Lednicky\thanks{%
e-mail: lednicky@fzu.cz }}
\address{Institute of Physics, Na Slovance 2,
\\18040 Prague 8, Czech Republic}

\maketitle

\begin{abstract}
Using the method of locally equilibrium statistical
operator we consider the thermalized relativistic quantum fields in an
oscillatory trap. We compare this thermal picture of the confined boson gas
with non-relativistic model of independent factorized sources. We find that
they are equivalent in the limit of very large effective sizes $R$, more
exactly, when the Compton wave length $1/m$ and thermal wave length $1/\sqrt{%
mT}$ are much smaller than {\ }$R$ . Under this conditions we study the
influence of Bose condensation in finite volumes on the structure of the
Wigner function, momentum spectra and 
correlation function.
\end{abstract}

%\draft

%\date{}
%\pacs{}

\section{Introduction}

In heavy ion collisions at RHIC and LHC energies quasi-macroscopical systems
containing $10^4-10^5$ particles are expected to be created. If phase-space
densities of such systems at a pre-decaying stage will be high enough, one
can observe multi-boson effects enhancing the production of pions with low
relative momenta, softening their spectra and modifying correlation
functions. One can even hope to observe new interesting phenomena like boson
condensation in certain kinematic regions with a large pion density in the
6-dimensional phase space: $f=(2\pi )^3d^6n/d^3{\bf p}d^3{\bf x}>\sim 1$
(see, e.g., \cite{zaj87,zha93,pra94,ber94,sl94}).

Generally, the account of the multi-boson effects is extremely difficult
task. Even on the neglection of particle interactions in the final state the
requirement of the BE symmetrization leads to severe numerical problems
which increase factorially with the number of produced bosons \cite
{zaj87,zha93}. In such situation, it is important that there exists a simple
analytically solvable models \cite{pra94} allowing for a study of the
characteristic features of the multi-boson systems. Actually, there are two
basic methods that are presently used to describe multi-boson systems in finite
(small) volumes typical for A+A collisions. First one \cite
{pra94,ame-led,era95} is maximally closed to the procedure of computer
simulations of multi-boson effects. We call it: the model of independent
factorized sources, or MIFS. It implies the independent emission of 'probed`
non-identical particles with factorized Wigner functions

\begin{equation}
D_n(p_1,x_{1;}...;p_n,x_n)=\prod\limits_{i=1}^nD(p_i,x_i).  \label{8.74a}
\end{equation}
The basic function in such an approach is the 'probed` Wigner distribution $%
D(p,x)$ that is chosen usually in the thermal-like (Boltzmann) form:

\begin{equation}
D(p,x)=\frac \eta {(2\pi R\Delta )^3}\exp (-\frac{{\bf p}^2}{2\Delta ^2}-%
\frac{{\bf r}^2}{2R^2})\delta (t).  \label{8.71}
\end{equation}
We normalize it to a mean multiplicity $\eta .$ In typical cases the input
numbers $n$ of particles are described by the Poissonian distribution

\begin{equation}  \label{poisson}
P(n)=e^{-\eta}\eta ^n/n!.
\end{equation}
In order to obtain an output ('true`) particle number distribution as well
as single and two-particle momentum spectra, the special procedure of
''switching on'' of the Bose-statistics is applied. In other words, the
symmetrization of the 'probe' $n$-particle normalized amplitudes $A_n\{p_i\}$
is provided for. Then the final multiplicity distribution is $%
P^S(n)=P(n)N[A_n^S]$ , where $N[A_n^S]$ is the normalized weight due to
the symmetrization.

The other approach \cite{cha95,zim97}deals with true boson statistics from
very beginning and is based on the density matrix formalism. In particular,
the specific $\rho -$matrix ansatz in the wave packet formalism was proposed 
\cite{zim97} to get (and develop) MIFS results.

To treat the results of these approaches one can often use the statistical
thermodynamics language like ''rare gas'', ''Bose condensate'', etc. At the
same time no systematic consideration based on the thermal matrix density
for such a kind of system in finite volumes has been done. For thermal
relativistic essentially finite (small) systems the problem is
mathematically rather complicated, however for large enough ones 
it can be solved
analytically. We will demonstrate it here using the method of
statistical operator.

\section{Density matrix for locally equilibrium systems}

According to the definition the statistical operator is

\begin{equation}
{\bf \rho =}e^{-S}  \label{Rho-S}.
\end{equation}
For locally equilibrium systems the entropy $S$ has to be maximized under
additional conditions fixing the average densities of energy $\epsilon (x)$,
momentum ${\bf p}(x)$, charge $q(x)$. Systems are considered on some
hypersurface $\sigma :d$$\sigma _\nu =d\sigma n_\nu $ with a time-like
normal vector $n^\nu $. In the relativistic covariant form the conditions
look like

\begin{equation}
p^\mu (x)\equiv \left( \epsilon (x),{\bf p}(x)\right) =\left\langle n_\nu (x)%
\widehat{T}^{\mu \nu }(x)\right\rangle ,\quad q(x)=\left\langle n_\nu (x)%
\widehat{J}^\nu (x)\right\rangle  \label{additional-cond}
\end{equation}
where $\left\langle ...\right\rangle $ means the average with 
the statistical operator ${\bf %
\rho }$ in (\ref{Rho-S}). The result for entropy operator is then \cite{Zub}, 
\cite{VonWeert}

\begin{equation}
S=S(\sigma )=\Phi (\sigma )+\int d\sigma _\nu (x)\beta _\mu (x)\widehat{T}%
^{\mu \nu }(x)-\int d\sigma _\nu (x)\mu (x)\widehat{J}^\nu (x),  \label{S}
\end{equation}
\noindent where $\Phi (\sigma )=\ln \;Sp\exp \{\int d\sigma \;n_\nu (x)\beta
_\mu (x)\widehat{T}^{\mu \nu }(x)\}$ is Masier-Planck functional, $\beta
_\mu (x)$ and $\mu (x)$ are Lagrange multipliers. For real one-component
free scalar field we will use the current of particle number density $%
\widehat{J}^\nu (x)=\varphi ^{(+)}(x)\stackrel{\longleftrightarrow }{%
\partial ^\nu }\varphi ^{(-)}(x)$ where $\varphi ^{(+)}$ and $\varphi ^{(-)}$
are the positive and negative field components. The energy-momentum tensor
has the standard form.

Then for the system which 
has no internal flows and
is in local equilibrium state on the hypersurface $%
t=0$: $n^\nu =(1,{\bf 0),\ }$ at the same temperature $%
T$ : $\beta _\mu (x)=(\frac 1T{\bf ,0)\ }$the density matrix looks like:

\begin{equation}
\rho =\frac 1Z\exp \left[ -\beta \int d^3pp^0a_p^{\dagger }a_p+\frac \beta
{2(2\pi )^3}\int d^3x\mu (x)\frac{d^3kd^3k^{\prime }}{k_0k_0^{\prime }}%
(k_0+k_0^{\prime })e^{-i{\bf (k-k}^{\prime }{\bf )x}}a_k^{\dagger
}a_{k^{\prime }}\right]   \label{Rho-our}
\end{equation}
To guarantee the maximal closeness to the MIFS we will use the chemical
potential in the form:

\begin{equation}  \label{mu(x)}
\mu (x)=\mu -\frac{{\bf x^2}}{2R^2\beta }.
\end{equation}

The description of inclusive spectra and correlations for a multiparticle
production is based on a computation of the averages

\begin{equation}
p^0\frac{dN}{d{\bf p}}=\left\langle a_p^{+}a_p\right\rangle ,\text{ }%
p_1^0p_2^0\frac{dN}{d{\bf p}_1d{\bf p}_2}=\left\langle
a_{p_1}^{+}a_{p_2}^{+}a_{p_1}a_{p_2}\right\rangle ,\text{ etc,...}
\label{spectra-def}
\end{equation}
\noindent  For Gaussian-type operator like (\ref{Rho-our}) the thermal Wick
theorem takes place:

\begin{equation}
\left\langle a_{p_1}^{+}a_{p_2}^{+}a_{p_1}a_{p_2}\right\rangle =\left\langle
a_{p_1}^{+}a_{p_1}\right\rangle \left\langle a_{p_2}^{+}a_{p_2}\right\rangle
+\left\langle a_{p_1}^{+}a_{p_2}\right\rangle \left\langle
a_{p_2}^{+}a_{p_1}\right\rangle . \label{wick}
\end{equation}
To find the averages $\left\langle a_{p_1}^{+}a_{p_2}\right\rangle $ the
Gaudin's method \cite{Gaudin} is used \cite{sin93}. As a result we have:

\begin{equation}
\left\langle a_{p_1}^{+}a_{p_2}\right\rangle =p_2^0\sum_{n=1}^\infty \
G_n(p_1,p_2);\text{ }G_n(p_1,p_2)=\int d^3kG_{n-1}(p_1,k)G_1(k,p_2).
\label{aa}
\end{equation}
The basic function $G_1(p_1,p_2)$ will be defined below. For this aim,
first the commutation relation with entropy operator have to be considered:

\begin{equation}
\lbrack a(p),S]=\int d^3kM(p,k)a(k)\Longrightarrow M(p,k)\approx
M^{(0)}(p,k)+O(\frac 1{p_0^2R^2})+O(\frac \beta{p_0R^2})  \label{M}
\end{equation}
where, using (\ref{Rho-our}) and (\ref{mu(x)}), we have

\begin{equation}
M^{(0)}(p_1,p_2)=\beta p_2^0\delta ({\bf p}_1-{\bf p}_2)-\frac \beta {(2\pi
)^3}\int d^3xe^{-i{\bf (p}_1{\bf -p}_2{\bf )x}}\mu (x).  \label{M0}
\end{equation}
The basic function $G_1(p_1,p_2)$ is then defined as follows:

\begin{equation}
G_1(p_1,p_2)=\sum_{n=0}^\infty \frac{(-1)^n}{n!}M_n^{*}(p_1,p_2),  \label{G}
\end{equation}
where

\begin{eqnarray}
M_0(p_{1,\text{ }}p_2) &=&\delta ({\bf p}_1-{\bf p}_2),\text{ }M_1(p_{1,%
\text{ }}p_2)=M(p_{1,\text{ }}p_2);  \nonumber  \label{Mn} \\
&&  \label{Mn} \\
M_n(p_1,p_2) &=&\int d^3kM_{n-1}(p_1,k)M_1(k,p_2).  \nonumber
\end{eqnarray}

\section{The comparison with MIFS.}

At $m^2R^2,m\beta ^{-1}R^2\gg 1$ we find, 
neglecting the terms $%
1/p_0^4R^4$ and $ \beta ^2/p_0^2R^4$ in Eqs.(\ref{M}),

\begin{equation}
\begin{array}{l}
G_1(p_1,p_2)=G_1^{(0)}(p_1,p_2)+G_1^{(1)}(p_1,p_2)+O(\frac 1{p_0^4R^4},\frac{%
\beta ^2}{p_0^2R^4})\approx  \\ 
\\ 
\frac 1{(2\pi )^3}\int d^3xe^{i{\bf qx}}\exp (-\beta (p_1^0-\mu (x)))\times 
\\ 
\\ 
\left[ 1-\frac{3\beta }{4p_1^0R^2}(1+\frac{2i{\bf p}_1{\bf x}}3+\frac{2\beta 
{\bf p}_1^2}{9p_1^0})-\frac 3{4(p_1^0R)^2}(1+\frac{\beta {\bf p}_1^2}{p_1^0}%
)\right] =\int d^3xe^{i{\bf qx}}D(x,p).
\end{array}
\label{G1}
\end{equation}
Here $p=(p_1+p_2)/2;$ $q=p_1-p_2.$ At $m^2R^2,\Delta ^2R^2\rightarrow $ $%
\infty $ in the non-relativistic limit the Wigner distribution is

\begin{equation}
D(x,p)\approx \frac \xi {(2\pi )^3}\exp (-\frac{{\bf p}^2}{2\Delta ^2}-\frac{%
{\bf r}^2}{2R^2}),  \label{D}
\end{equation}
where $\Delta ^2=mT$ and fugacity $\xi =e^{\widetilde{\mu }},\widetilde{\mu }%
=\mu -m.$ Comparing Eqs. (\ref{D}) and (\ref{8.71}) we found that in the
non-relativistic approach in the limit of very large emission volumes the
thermal density matrix (\ref{Rho-our}) with the
chemical potential (\ref{mu(x)}) and the fugacity $\xi
=\eta /(\Delta R$ $)^3$ is equivalent to MIFS. The physical reasons are the
following. First, if the wave-length of the quanta is much less that the
system size, the particle does not 'feel' the finiteness of the systems and
the 'probe' distribution will be the same as in the thermodynamic limit,
i.e., the Boltzmann-like one. Second, for large systems the assumption of an
independent emission of non-interacting particles is natural: the average
distance between any two particles is large comparing with its wave-length
(or with the size of the wave packet).

At the same time Eq.(\ref{G1}) indicates that for essentially small system's
sizes compared with quanta wave-lengths, the single-particle locally
equilibrium distribution $D(x,p)$ has no more simple Boltzmann-like form (%
\ref{8.71}). Some distortion terms have relativistic nature and are
essential when the system size is close to the Compton wave-length, $R\sim
1/m.$ The others can take place even if $Rm\gg 1$, depending on the ratio
between size $R$ and thermal wave-length $1/\sqrt{mT}$. They are of
quantum nature and appear when one describes the thermal equilibrium of the
quanta with an average de Broglie wave length that is larger than the system
size. In general, according to Eq.(\ref{G1}) one can expect the reduction of
soft quanta in thermal model in comparison with the MIFS Boltzmann anzats
while the distributions of ''hard'' quanta  coincide in  both
approaches.

\section{The Wigner functions and spectra in multi-boson systems}

We will consider the limiting behavior 
of the functions $G_n({\bf p}%
_1,{\bf p}_2)$ (\ref{aa}) at small $n<<\Delta R$ and large $n\gg \Delta R$
at $R\Delta \gg 1$ in the non-relativistic case. Neglecting the terms 
  $1/\Delta ^2R^2$ and $ 1/m^2R^2$ in Eq. (\ref{G1}) we
can put $G_1(p_1,p_2)=G_1^{(0)}(p_1,p_2)$ and found the optimal tailing of the 
two limiting forms at the point $n_t=R\Delta .$ Then at $R\Delta \gg 1$
one can express the basic operator average (\ref{aa}) through the Wigner
functions of the Bose gas (g) and the Bose condensate (c) (we will see the
correspondence later) as follows:

\begin{equation}
\frac 1{p_2^0}\left\langle a_{p_1}^{+}a_{p_2}\right\rangle
=\sum_{n=1}^\infty G_n(p_1,p_2)\approx \int d^3xe^{i{\bf qx}%
}(f_g(p,x)+f_c(p,x)),  \label{wigner}
\end{equation}
where

\begin{eqnarray}
f_g(p,x) &=&\frac 1{(2\pi )^3}\frac 1{\widetilde{\xi }_x^{-1}\exp ({\bf p}%
^2/2\Delta ^2)-1}\text{ },  \nonumber \\
&&  \label{wigner_g} \\
\widetilde{\xi }_x &=&\widetilde{\xi }\exp (-{\bf x}^2/2R^2)\equiv e^{\beta (%
\widetilde{\mu }-\frac 3{2\Delta R})}\exp (-{\bf x}^2/2R^2)  \nonumber
\end{eqnarray}
and

\begin{equation}
f_c(p,x)=\frac{\widetilde{\xi }}{1-\widetilde{\xi }}\widetilde{\xi }%
^{R\Delta }\frac 1{\pi ^3}\exp (-\frac R\Delta {\bf p}^2-\frac \Delta R{\bf x%
}^2).  \label{wigner_c}
\end{equation}
We have used here the following transformation:

\begin{equation}
\exp \left( -\frac{{\bf q}^2R^2}{2n}\right) =\left( \frac n{2\pi R^2}\right)
^{3/2}\int d^3xe^{-i{\bf xq}}\exp (-\frac{{\bf x}^2n}{2R^2}).  \label{transf}
\end{equation}

Note here that the critical value of the nonrelativistic chemical potential $%
\widetilde{\mu }=\frac 3{2\Delta R}$ is positive because gas is not ideal
(effectively the gas interacts with external field confining it in a finite
volume).

In the thermodynamic limit: volume $V\rightarrow \infty $ ($R\Delta
\rightarrow \infty )$ at fixed temperature $T$ and at fixed particle density 
$n=N/V$ the phase space density in the central part is 
\begin{equation}
\widetilde{n}\equiv \frac{n(0)}{\Delta ^3}\approx \frac 1{(2\pi )^{3/2}}\Phi
_{3/2}(\widetilde{\xi })+\frac 1{(\pi R\Delta )^{3/2}}\frac{\widetilde{\xi }%
}{1-\widetilde{\xi }}\widetilde{\xi }^{R\Delta }\equiv \widetilde{n}_g+%
\widetilde{n}_c  \label{den-cen}
\end{equation}
where $\Phi _\alpha (z)=\sum_{k=1}^\infty \frac{z^k}{k^\alpha }$ . The first
term corresponds to the ordinary Bose-Einstein gas contribution. Let us
introduce associated with it at $\widetilde{\xi }=1$ the critical density: $%
\widetilde{n}_{cr}=\frac 1{(2\pi )^{3/2}}\Phi _{3/2}(1)$. It is easy to see
that in order to move to the thermodynamic limit at fixed density $%
\widetilde{n}>\widetilde{n}_{cr}$ the parameter $\widetilde{\xi }$ at large
enough $R\Delta $ has to tend to unity as:

\begin{equation}
\widetilde{\xi }=1-\frac 1{(\widetilde{n}-\widetilde{n}_{cr})(\pi R\Delta
)^{3/2}}.  \label{low}
\end{equation}

The single particle inclusive distribution in the thermodynamic limit has
the form:

\begin{equation}
\begin{array}{c}
{\large n^{(1)}}({\bf p})=\int\limits_{-\infty }^\infty d^3xf({\bf x,p)}=%
{\large n}_g({\bf p})+{\large n}_c({\bf p})\approx \\ 
\\ 
\frac{R^3}{(2\pi )^{3/2}}\Phi _{3/2}(\widetilde{\xi }e^{-\frac{{\bf p}2}{%
2\Delta ^2}})+\frac{\widetilde{\xi }}{1-\widetilde{\xi }}\widetilde{\xi }%
^{R\Delta }R^3(\pi R\Delta )^{-3/2}\exp (-\frac R\Delta {\bf p}^2) \\ 
\\ 
\stackrel{R\Delta \rightarrow \infty }{\longrightarrow }\frac 1{(2\pi
)^3}\int d^3x\frac 1{\widetilde{\xi }_x^{-1}\exp ({\bf p} ^2/2\Delta
^2)-1}+(\pi R\Delta )^{3/2}(\widetilde{n}-\widetilde{n} _{cr})\delta ({\bf p}%
)
\end{array}
\label{limspectra}
\end{equation}
Two terms in the last equality of Eq. (\ref{limspectra}) are associated with
the classical Bose gas and the condensate in thermodynamic limit at the
densities $\widetilde{n}>\widetilde{n}_{cr}$. They
describe Bose gas and Bose condensate for finite systems at the condition $%
R\Delta >>1$. Note that for Bose condensate the momentum distribution is
much more narrow than for Bose gas, $\sqrt{\frac \Delta{2R }}$ $\ll \Delta .$

\section{The interferometry of multi-boson sources}

The calculation of two-particle inclusive spectra is based on Eqs.(\ref
{spectra-def}), (\ref{wick}), (\ref{wigner}). One can easily find that

\begin{equation}
\frac 1{p_2^0}\left\langle a_{p_1}^{+}a_{p_2}\right\rangle ={\large n}_g(%
{\bf p})\exp (-{\bf q}^2R_g^2/2)+{\large n}_c({\bf p})\exp (-{\bf q}%
^2R_c^2/2),  \label{average}
\end{equation}
where 
\begin{equation}
R_g^2=R^2\Phi _{5/2}(\widetilde{\xi }e^{-\frac{p2}{2\Delta ^2}})/\Phi _{3/2}(%
\widetilde{\xi }e^{-\frac{p2}{2\Delta ^2}})\stackrel{p=0,\widetilde{\xi }=1}{%
\longrightarrow }\approx R^2/2,\text{ }R_c^2=R/2\Delta .  \label{RR}
\end{equation}
The correlation function at small $q^2$ limit, $qR\ll 1,$ is then: 
\begin{equation}
C(p,q)=1+\frac{\left\langle a_{p_1}^{+}a_{p_2}\right\rangle \left\langle
a_{p_2}^{+}a_{p_1}\right\rangle }{\left\langle
a_{p_1}^{+}a_{p_1}\right\rangle \left\langle a_{p_2}^{+}a_{p_2}\right\rangle 
}=1+\exp \left( -\frac{{\large n}_g({\bf p})}{{\large n}_g({\bf p})+{\large n%
}_c({\bf p})}R_g^2{\bf q}^2\right).   \label{q-small}
\end{equation}
At large $q^2$ limit, $qR\gg 1$, it is 
\begin{equation}
C(p,q)=1+\left( \frac{n_c}{n_g}\right) ^2\exp (-R_c^2q^2).  \label{q-large}
\end{equation}
The effective interferometry radius squared, 
corresponding to $C(q_{eff})=1+1/e$, is

\begin{equation}
R_{eff}^2\equiv q_{eff}^{-2}=R_c^2/\left[ 1+2\ln (n_c(p)/n_g(p))\right].
\label{Reff}
\end{equation}

Finally, we can see that when the density in phase space increases the
interferometry radius of the gas component reduces to $R/\sqrt{2}$ at
most (see (\ref{RR})). At very large densities, exceeding essentially the
critical one, the Bose condensate  determines the correlation function
behavior at small $p.$ Then the interferometry radius is reduced
additionally as compared with its behavior for pure Bose-gas. At $\widetilde{%
n}/\widetilde{n}_{cr}\rightarrow \infty ,$ the interferometry radius in
inclusive measurements tends to zero whatever large is the geometric size of
the system. Note that intercept of the inclusive correlation function is
equal to 2 and does not depend on $\widetilde{n}/\widetilde{n}_{cr}$. This
reflects the chaotic nature of the thermal multiboson source. We are not
discussing here the very complicated problem of spontaneous symmetry breaking
of the condensate in finite systems which could take place when there is the
interaction destroying the degeneration in the system.

\end{document}